\documentstyle[12pt]{article}
\begin{document}

\baselineskip=23pt

\begin{center}
{\Large\bf Global geometric properties of AdS space\\
and the AdS/CFT correspondence}

\bigskip

\bigskip

Qi-Keng LU


{\em Institute of Mathematics\\
Academy of Mathematics and Systems Sciences, Academia Sinica\\
Beijing 100080, China\footnote{Email: luqik@public.bta.net.cn.}\\}

\medskip

Zhe CHANG

{\em
Institute of High Energy Physics, Academia Sinica\\
P.O.Box 918(4), Beijing 100039, China\footnote{
 Email: changz@alpha02.ihep.ac.cn. }\\}

\medskip

 Han-Ying GUO


{\em Institute of Theoretical Physics, Academia Sinica\\
P.O.Box 2735, Beijing 100080, China\footnote{ Email: hyguo@itp.ac.cn.}}

\bigskip

\end{center}

\bigskip

{\bf Abstract} $~~~~$ The Poisson kernels  and relations between them
for a massive scalar field in a unit ball $B^{n}$ with Hua's
metric and conformal flat metric are obtained by describing the $B^{n}$ as
a submanifold of an $(n+1)$-dimensional embedding space. Global geometric
properties of the AdS space are discussed. We show that the
$(n+1)$-dimensional AdS space AdS$_{n+1}$ is isomorphic to $RP^1\times B^n$
and boundary of the AdS is isomorphic to $RP^1\times S^{n-1}$.
Bulk-boundary propagator and the AdS/CFT like correspondence are demonstrated
based on these global geometric properties of the $RP^1\times B^n$.\\ \\
{\bf Keywords: Poisson kernel, AdS/CFT correspondence, Bulk-boundary\\
$~~~~~~~~~~~~~~~~~$propagator}

\newpage

\section{Introduction}
In Riemann geometry, it is known that the biggest
symmetry(isometric) groups of the Minkowski space, de Sitter Space
(dS) and Anti-de Sitter (AdS) space have same number of
generators. They can be identified, in a unified way, as the
classical manifold $D_\lambda(n+1)$ with $\lambda=0,~+1$ and $-1$,
respectively. Thus, the dS and the AdS space are the simplest
generalization of the Minkowski space with constant curvature. The
dS,  in particularly, the AdS space and quantum field theory based
on it has been an interested topic of mathematicians and
physicists for a long time\cite{01}. There has recently been a
revival of interest in AdS space brought about by the conjectured
duality between physics in the bulk of AdS and a conformal field
theory (CFT) on the boundary\cite{02}--\cite{04}. The so-called
AdS/CFT correspondence states that string theory in the AdS space
is holographically dual to a CFT on boundary of the AdS\cite{05}.
A strong support for the proposal comes from comparing spectra of
Type IIB string theory on the background of AdS$_5\times S^5$ and
low-order correlation functions of the $3+1$ dimensional ${\cal
N}=4$ $SU(N)$ super Yang-Mills theory. The dual super Yang-Mills
theory lives on the boundary of the AdS space. This is one of the
most important progresses in the superstring theory. Many new
results have been obtained by making use this conjecture\cite{06}.

However, up to now, almost all discussions on the AdS/CFT correspondence
were based on the so-called Euclidean version of the AdS, or an
$(n+1)$-dimensional unit ball $B^{n+1}$. To prove the AdS/CFT conjecture or
to investigate its delicious implications in physics theory, one should
work on the more challenging topic of duality between physics theories on
the AdS space and its boundary, the compact Minkowski space.

In this paper, by describing the unit ball $B^{n}$ as a submanifold of an
$(n+1)$-dimensional embedding space, we first present Poisson kernels for
the Laplace operator with a general nonzero  eigenvalue $m_0$ and relations
between them
in a unit ball $B^{n}$ with Hua's metric and conformal flat metric. Results
for Euclidean version of the AdS are recovered in a different view. Then we
discuss the AdS geometry with right signature within the framework of
the classical manifolds and
the classical domains\cite{07,08}. It is proved that the AdS$_{n+1}$ is isomorphic
to $RP^1\times B^n$ and the boundary of the AdS is isomorphic to $RP^1\times
S^{n-1}$. Bulk-boundary propagator for a massive scalar field in
the AdS is replaced by what in the $RP^1\times B^n$ based on the global geometric properties of
AdS. The bulk/boundary correspondence in this  case is also demonstrated.

\section{Poisson kernels for a scalar field in unit ball}

A unit ball $B^{n}$ can be described as the image of a two-to-one
map of the hypersurface

\begin{equation}
\xi^{n}\xi^{n}-\sum_{i=0}^{n-1}\xi^i\xi^i=1
\end{equation}
in the space $(\xi^0, \xi^1,\cdots, \xi^{n-1})$.

 An explicitly $SO(1,n)$ invariant metric of the $B^n$ can be
introduced
\begin{equation}\label{metric}
ds^2=d\xi^n d\xi^n- \sum_{i=0}^{n-1}d\xi^id\xi^i~.
\end{equation}
By denoting
\begin{equation}
x^i=\frac{\xi^i}{\xi^{n}}~,~~~~\left(i=0,~1,~\cdots,~n-1~,~~~\xi^{n}\not= 0
\right)~,
\end{equation}
$n$-dimensional vector $x\equiv (x^0,x^1,\cdots,x^{n-1})$, and $x'$ transfer
of the $x$, we can rewrite the $B^n$ as usual
\begin{equation}
xx'<1~.
\end{equation}
The reduced metric from Eq.(\ref{metric}) in the coordinate $\{x^i\}$ is of the form
\begin{equation}\label{dg01}
ds^2=-\frac{dx(I-x'x)^{-1}dx'}{1-xx'}~.
\end{equation}

Another set of coordinates for the unit ball $B^n$ can be introduced
\begin{equation}
z^i=\frac{\xi^i}{\xi^{n}+1}~,~~~~\left(i=0,~1,~\cdots,~n-1~;~~~\xi^{n}
\not=0\right)~.
\end{equation}
In this coordinate, the unit ball $B^n$ is also described as usual
\begin{equation}
zz'<1~.
\end{equation}
The reduced metric from Eq.(\ref{metric}) in terms of the coordinate
$\{z^i\}$ is
\begin{equation}\label{dg02}
ds^2=\frac{-4}{(1-zz')^2}dz dz'~.
\end{equation}
There is a one to one transformation between the two sets of coordinates $\{x^i\}$
and $\{z^i\}$
\begin{equation}\label{tran}
x^i=\frac{2z^i}{1+zz'}~.
\end{equation}
Therefore, the conformal flat metric (\ref{dg02}) and the Hua's
metric (\ref{dg01}) are different representations of the $SO(1,n)$
invariant metric. And we can work by using one of them and got same
results include invariant differential form and Poisson kernel.

The equation of eigenvalues for the Laplace operator in the unit ball $B^n$ is
\begin{equation}
\left[\frac{1}{\sqrt{-g}}\sum_{i,j=0}^n\frac{\partial}{\partial
x^i}\left(\sqrt{-g}g^{ij} \frac{\partial}{\partial
x^i}\right)+m_0^2\right]\Phi(x)=0~.
\end{equation}
The Poisson kernel for a massless scalar field has been discussed\cite{09}
\begin{equation}
G^E_{B\partial}(x,u)=\left\{\begin{array}{l}
                            \displaystyle\frac{(1-xx')^{n-1}}{(1-2ux'+
                            xx')^{n-1}}~,~~~~
                            {\rm for ~conformal ~flat ~metric~,}\\[1cm]
                            \displaystyle\frac{(1-xx')^{\frac{n-1}{2}}}
                            {(1-ux')^{n-1}}~,~~~~{\rm for ~Hua's ~metric}~.
                            \end{array}\right.
\end{equation}
The bulk field $\Phi(x)$ determined by the fields living on the boundary
$\phi(u)$ is of the form
\begin{equation}
\Phi(x)=\frac{1}{\omega_{n-1}}\int_{uu'=1}\cdots\int G^E_{B\partial}
(x,u)\phi(u)\dot{u}~.
\end{equation}
Now, we write down a general Poisson kernel for the Laplace operator
with nonzero eigenvalue $m_0^2$
\begin{equation}
G^\pm_{B\partial}({x},u)=\left(G^E_{B\partial}({x},u)
\right)^{\frac{1}{2}\left(1\pm\sqrt{1+\frac{4m_0^2}
{(n-1)^2}}\right)}~.
\end{equation}
The Poisson kernels $G^\pm_{B\partial}({x},u)$ satisfy the
following properties:
\begin{itemize}
\item It is definitely positive.
\item On the boundary, we have
      \begin{equation}
      G^\pm_{B\partial}(v,u)=\left\{\begin{array}{c}
                                           0~,~~~~~u\not=v~,\\
                                           \infty~,~~~~~u=v~.
                                           \end{array}\right.
      \end{equation}
\item It satisfies the equation of eigenvalues for the Laplace operator with
a eigenvalue $m_0^2$.
\end{itemize}

\section{Conformal boundary and AdS}

In an $(n+2)$-dimensional embedding space, the $(n+1)$-dimensional AdS space
AdS$_{n+1}$ can be written as
 \begin{equation}\label{qiu}
  \xi^0\xi^0-\sum_{i=1}^n\xi^i\xi^i+\xi^{n+1}\xi^{n+1}=1~.
 \end{equation}
From the above definition of AdS, we know that $\xi^0$ and $\xi^{n+1}$ can not be
zero simultaneously, and at least two charts of coordinates
 $[(n+1)$-dimensional] ${\cal U}_1$ ($\xi^{n+1}\not= 0$) and
 ${\cal U}_0$ ($\xi^0\not= 0$) should be
needed to describe the AdS.

In the chart AdS$_{n+1}\cap {\cal U}_1$, we introduce a coordinate
 \begin{equation}
  x^i=\frac{\xi^i}{\xi^{n+1}}~,~~~~~~(i=0,~1,~2,~\cdots,~n;~~\xi^{n+1}\not=0)~.
 \end{equation}
The AdS$_{n+1}\cap {\cal U}_1$, in the coordinate $\{x^i\}$, is described by
\begin{equation}
\begin{array}{l}
{\rm AdS}_{n+1}\cap {\cal U}_1:~~~~~~\sigma(x^i,x^j)>0~,\\[0.5cm]
\sigma(x^i,x^j)\equiv 1+\displaystyle\sum_{i,j=0}^n\eta_{ij}x^i x^j ~,~~~~~\eta={\rm diag}
(1,~\underbrace{-1,~-1,~\cdots,~-1}\limits_{n})~.
\end{array}
\end{equation}
In the chart AdS$_{n+1}\cap {\cal U}_0$, let
\begin{equation}
\begin{array}{l}
\displaystyle y^0=\frac{\xi^{n+1}}{\xi^0}~,\\[0.5cm]
\displaystyle y^i=\frac{\xi^{i}}{\xi^{0}}~,~~~~~~(i=1,~2,~\cdots,~n;~~
\xi^{0}\not=0)~.
\end{array}
\end{equation}
And the AdS$_{n+1}\cap {\cal U}_0$ can be written in the form
\begin{equation}
{\rm AdS}_{n+1}\cap {\cal U}_0:~~~~~~\sigma(y^i,y^j)>0~.
\end{equation}
At the overlap region AdS$_{n+1}\cap{\cal U}_0\cap {\cal U}_1$ of the two
charts ${\cal U}_1$ and ${\cal U}_0$, one
has relations
\begin{equation}
\begin{array}{l}
\displaystyle y^0=\frac{1}{x^0}~,\\[0.5cm]
\displaystyle y^i=\frac{x^i}{x^0}~.
\end{array}
\end{equation}
This shows clearly a differential structure\cite{10} of the AdS.
The boundary $\overline{M}^n$ of the AdS$_{n+1}$ consists of infinite
points not belong to AdS$_{n+1}\cap{\cal U}_1$ or AdS$_{n+1}\cap{\cal U}_0$,
\begin{equation}
\begin{array}{r}
\overline{M}^n:~~1+\displaystyle\sum_{i,j=0}^n\eta_{jk}x^jx^k=0~,\\[0.5cm]
1+\displaystyle\sum_{i,j=0}^n\eta_{jk}y^jy^k=0~.
\end{array}
\end{equation}

A new set of variables in the chart AdS$_{n+1}\cap {\cal U}_1$ can be
introduced as
\begin{equation}
\begin{array}{l}
\chi^0\equiv x^0~,\\
\chi^\mu\equiv\sqrt{1+(x^0)^2}x^\mu~,~~~(\mu=1,~2,~\cdots,~n).
\end{array}
\end{equation}
Then, we have
\begin{equation}
{\rm AdS}_{n+1}\cap {\cal U}_1=\{\chi^0\in \Re,~(\chi^1,~\chi^2,~\cdots,~
\cdots,~\chi^n)\in \Re^n\vert(\chi^1)^2+(\chi^2)^2+\cdots+(\chi^n)^2<1\}~.
\end{equation}
This shows that, in the chart ${\cal U}_1$,  the AdS$_{n+1}$ is equivalent to
$\Re\times B^n$.

In the same way, in the chart AdS$_{n+1}\cap{\cal U}_0$, let
 \begin{equation}
\begin{array}{l}
\eta^0\equiv y^0~,\\
\eta^\mu\equiv\sqrt{1+(y^0)^2}y^\mu~,~~~(\mu=1,~2,~\cdots,~n).
\end{array}
\end{equation}
And subsequently, one has
 \begin{equation}
{\rm AdS}_{n+1}\cap {\cal U}_0=\{\eta^0\in \Re,~(\eta^1,~\eta^2,~\cdots,~
\cdots,~\eta^n)\in \Re^n\vert(\eta^1)^2+(\eta^2)^2+\cdots+(\eta^n)^2<1\}~.
\end{equation}
Therefore, both charts of the AdS$_{n+1}$, ${\cal U}_1$ and ${\cal U}_0$ are
equivalent to $\Re\times B^n$.

 It should be noticed that, at the overlap region of the two charts,
there are relations among the two different sets of coordinate variables
\begin{equation}
\begin{array}{l}
\displaystyle \chi^0=\frac{1}{\eta^0}~,\\
\chi^\mu=\eta^\mu~,~~~(\mu=1,~2,~\cdots,~n).
\end{array}
\end{equation}
This fact presents a theorem for the AdS.\\
{\bf Theorem}: {\em The AdS$_{n+1}$ is isomorphic to $RP^1\times B^n$ and
its  boundary  is $RP^1\times S^{n-1}$.}.

It is well-known that the Study-Fubini metric can be introduced on the
$RP^1$ space
\begin{equation}
ds_0^2=\frac{(d\chi^0)^2}{[1+(\chi^0)^2]^2}=(d\arctan\chi^0)^2~.
\end{equation}
As presented at the previous section, on the unit ball $B^n$, we have the Hua's
metric
\begin{equation}
ds_n^2=-\sum_{\mu,\nu=1}^nd\chi^\mu d\chi^\nu\left(\frac{\delta_{\mu\nu}}
{1-\displaystyle\sum_{\alpha=1}
^n\chi^\alpha\chi^\alpha}+\frac{\chi^\mu\chi^\nu}{(1-
\displaystyle\sum_{\alpha=1}^n
\chi^\alpha\chi^\alpha)^2}\right)~.
\end{equation}
Thus, a natural metric on the $RP^1\times B^n$ is of the form
\begin{equation}\label{fubini}
\begin{array}{rcl}
ds^2&=&ds_0^2-ds_n^2\\
    &=&\displaystyle\frac{(d\chi^0)^2}{[1+(\chi^0)^2]^2}-
       \sum_{\mu,\nu=1}^n d\chi^\mu d\chi^\nu
       \left(\frac{\delta_{\mu\nu}}{1-\displaystyle\sum_{\alpha=1}
^n\chi^\alpha\chi^\alpha}+\frac{\chi^\mu\chi^\nu}{(1-\displaystyle\sum_{\alpha=1}^n
\chi^\alpha\chi^\alpha)^2}\right)~.
\end{array}
\end{equation}
But, with this metric the $RP^1\times B^n$  is no longer AdS group invariant.
In what follows, we discuss the bulk/boundary correspondence and related
topics in this case.

\section{Eigenfunctions of Laplace operator on $RP^1\times B^n$}

Let

\begin{equation}
\theta={\rm arctan}\chi^0~.
\end{equation}
Then ${ RP}^1\times{B}^n$ and ${RP}^1\times{S}^{n-1}$ are
isomorphic to $S^1\times B^n$ and $S^1\times S^{n-1}$
respectively. Moreover,

\begin{equation}
\Box=\frac{\partial^2}{\partial\theta^2}-\Delta ~,
\end{equation}
where

\begin{equation}
\Delta=\frac{1}{\sqrt g}\sum_{i,j=1}^{n}\frac{\partial}{\partial
}{\partial\chi^i}\Big(\sqrt g g^{ij}\frac{\partial}{\partial
\chi^j}\Big)
\end{equation}
is the Laplace-Beltrami operator of the ball $B^n$.

Denote

\begin{equation}
\alpha_k(m_0)=1+\left[1+4\frac{m_0^2-k^2}{(n-1)^2}\right]^{\frac{1}{2}}
\end{equation}
and

\begin{equation}
P_{\alpha_k(m_0^2)}(\chi,u)=\left[G_{B\partial}(\chi,u)\right]^
{\alpha_k(m_0)}~,
\end{equation}
where $\chi=(\chi^1,\cdots,\chi^n)$. It can be proved that

\begin{equation}
\Delta P_{\alpha_k(m_0)}(\chi,u)=-(k^2-m_0^2)P_{\alpha_k(m_0)}(\chi,u)~.
\end{equation}

\section{Bulk-boundary propagator on $S^1\times B^n$ }

Let $\Phi_0(\varphi,u)$ be a field on the boundary $S^1\times S^{n-1}$
of $S^1\times B^n$. Develop it into Fourier series of $\varphi$ such
that

\begin{equation}
\Phi_0(\varphi,u)=\sum_{k=0}^{\infty}\left[a_k(u)\cos k\varphi
+b_k(u)\sin k\varphi\right]~,
\end{equation}
where
\begin{equation}
a_k(u)=\frac{1}{2\pi}\int_0^\infty\Phi_0(\varphi,u) \cos k\varphi~d\varphi~,
~~~b_k(u)=\frac{1}{2\pi}\int_0^{2\pi}\Phi_0(\varphi,u)
\sin k\varphi~d\varphi~.
\end{equation}
Construct a scalar field $\Phi(\theta,\chi)$ on $S^1\times B^n$ such that

\begin{equation}
\Phi(\theta,\chi)=\sum_{k=0}^{\infty}\left[\phi_k(\chi)\cos k\theta+
\psi_k(\chi)\sin k\theta\right]~,
\end{equation}
where

\begin{equation}
\begin{array}{l}
\phi_k(\chi)=\displaystyle\frac{1}{\omega_{n-1}}\int_{uu'=1}a_k(u)P_{\alpha_k(m_0)}
(\chi,u)\dot{u}~,\\
\psi_k(\chi)=\displaystyle\frac{1}{2\pi}\int_{uu'=1}b_k(u)
P_{\alpha_k(m_0)}(\chi,u)\dot{u}~.
\end{array}
\end{equation}
Since

\begin{equation}
\Delta\phi_k(\chi)=-(k^2-m_0^2)\phi_k(\chi)~~{\rm
and}~~\Delta\psi_k(\chi)=-(k^2-m_0^2)\psi_k(\chi)~,
\end{equation}
then $\Phi(\theta,\chi)$ must satisfy the equation

\begin{equation}
\Box\Phi(\theta,\chi)=-m_0^2\Phi(\theta,\chi)~.
\end{equation}
Finally, $\Phi(\theta,\chi)$ can be expressed into the form

\begin{equation}
\begin{array}{l}
\Phi(\theta,\chi)=\displaystyle\frac{1}{2\pi\omega_{n-1}}\sum_{k=0}^{\infty}
\int_{uu'=1}\left[a_k(u)\cos k\theta+b_k(u) \sin k\theta\right]
P_{\alpha_k(m_0)}(\chi,u)\dot{u}\\[0.5cm]
=\displaystyle\frac{1}{2\pi\omega_{n-1}}\sum_{k=0}^{\infty}\int_{uu'=1}
\int_{0}^{2\pi}\left[\Phi_0(\varphi,u)\cos k\varphi \cos k\theta
+\Phi_0(\varphi,u)\sin k\varphi\sin k\theta\right]P_{\alpha_k(m_0)}
(\chi,u)d\varphi \dot{u}~,
\end{array}
\end{equation}
or

\begin{equation}
\Phi(\theta,\chi)=\frac{1}{V(S^1\times S^{n-1})}\int_{S^1\times S^{n-1}}
\Phi_0(\varphi,u)G_{m_0}(\theta,\chi;\varphi,u)d\varphi\dot{u}~,
\end{equation}
where

\begin{equation}
G_{m_0}(\theta,\chi;\varphi,u)=\sum_{k=0}^{\infty}P_{\alpha_k(m_0)}
(\chi,u)\cos k(\varphi-\theta)
\end{equation}
is the propagator.

\vskip 2mm

\centerline{\bf Acknowledgments}

Two of us (Z.C. and H.Y.G. ) would like to thank S. K. Wang and K. Wu for
enlightening discussions.
The work was supported in part by the National Science Foundation of China.

\end{document}